\documentstyle[aps,version2,preprint]{revtex}    
\begin{document}

\draft

\begin{title}
Magnetic-field and chemical-potential effects on the low-energy separation 
\end{title}

\author{J. M. P. Carmelo$^{1}$ and A. H. Castro Neto$^{2}$}
\begin{instit}
$^{1}$ Department of Physics, University of \'Evora,
Apartado 94, P-7001 \'Evora Codex, Portugal\\
and Centro de F\'{\i}sica das Interac\c{c}\~oes Fundamentais,
I.S.T., P-1096 Lisboa Codex, Portugal
\end{instit}
\begin{instit}
$^{2}$ Department of Physics, University of California,
Riverside, CA 92521
\end{instit}
\receipt{May 1996}

\begin{abstract}
We show that in the presence of a magnetic field the
usual low-energy separation of the Hubbard chain is
replaced by a ``$c$'' and ``$s$'' separation. Here
$c$ and $s$ refer to small-momentum and low-energy 
independent excitation modes which couple both to charge and 
spin. Importantly, we find the exact generators of these 
excitations both in the electronic and pseudoparticle basis. 
In the limit of zero magnetic field these generators become the 
usual charge and spin fluctuation operators. The $c$ and $s$ elementary 
excitations are associated with the $c$ and $s$ pseudoparticles,
respectively. We also study the separate pseudoparticle left
and right conservation laws. In the presence of the magnetic
field the small-momentum and low-energy excitations
can be bosonized. However, the suitable bosonization
corresponds to the $c$ and $s$ pseudoparticle modes
and not to the usual charge and spin fluctuations.
We evaluate exactly the commutator between the electronic-density 
operators. Its spin-dependent factor is in general non diagonal 
and depends on the interaction. The associate bosonic
commutation relations characterize the present unconventional
low-energy separation.
\end{abstract}
\renewcommand{\baselinestretch}{1.656}   


\narrowtext

\section{INTRODUCTION}

One of the central properties of one-dimensional many-electron
problems \cite{Solyom} at zero magnetic field is the low-energy 
charge - spin separation \cite{Castro}. This is related to the 
vanishing of the one-electron renormalization factor in these 
Luttinger liquids \cite{Haldane,Anderson}. However, in contrast to the 
case of the ``g-ology'' and Luttinger models \cite{Solyom,Castro,Haldane}, 
the study of the Bethe-ansatz (BA) solvable models 
\cite{Bethe,Yang,Lieb,Korepinrev} did not include a
description of the low-energy physics in terms of operators
with known expressions in the usual electronic basis: 
the Luttinger-liquid character of these models relied on 
the identification of a structure in the low-energy spectrum provided 
by the BA \cite{Haldane,Schulz}. 

Further, although there has been considerable progress in 
understanding the critical-energy spectrum of multicomponent 
integrable systems in a magnetic field by combining BA and 
conformal-field theory \cite{Frahm}, this approach also does 
not provide an operator description of the low-energy problem. 
A $c$ and $s$ pseudoparticle operator representation of such
conformal-field theory of BA solvable models was introduced in Refs. 
\cite{Carmelo94,Carmelo94b}. However, the expressions of the 
corresponding pseudoparticle generators in terms
of the usual electronic operators remained an open question.
Therefore, this latter study was unable to characterize the
BA $c$ and $s$ excitation modes in terms of the usual charge or 
spin fluctuation operators. 

In this paper we follow the preliminary results of Ref. 
\cite{Carmelo94a} and derive the expressions of the low-energy 
and small-momentum $c$ and $s$ pseudoparticle generators 
\cite{Carmelo94,Carmelo94b} in terms of charge and spin 
electronic operators. This allows the description of these 
generators in terms of the usual charge and
spin fluctuation operators for different values of the
magnetic field and chemical potential. By writing such
generators explicitly in the electronic basis we show that 
$c$ and $s$ are not in general charge and spin, respectively: 
for finite values of the magnetic field and chemical
potential the $c$ and $s$ pseudoparticle-pseudohole modes
couple both to the charge and spin degrees of freedom. 
However, in the limits of zero magnetic field and zero chemical 
potential the $c$ and $s$ generators become the usual 
charge and spin fluctuation operators, respectively.

The low-energy physics of the present many-electron 
problem can be described in terms of a Landau liquid of 
pseudoparticles. Such Landau-liquid approach was introduced in Refs. 
\cite{Carmelo91,Carmelo91b,Carmelo92a,Carmelo92,Carmelo92b,Carmelo93}.
We show that this does not contradict its Luttinger-liquid
character \cite{Haldane,Carmelo94a} which is
explained by our pseudoparticle approach.
We generalize for BA solvable many-electron
problems the construction of the low-energy physics
in terms of separate left and right conservation laws.

We also study the pseudoparticle $c$ and $s$ 
bosonic-operator algebra which describes the low-energy 
physics of integrable electronic Luttinger liquids in a 
magnetic field. In contrast to the bosonization
scheme of the Luntinger model \cite{Solyom,Haldane},
the present bosonization refers to the many-electron
ground state. In terms of the pseudoparticle 
interactions the latter has a non-interacting character --
it is a simple Slatter determinant of pseudoparticle
levels \cite{Carmelo94,Carmelo94a,Carmelo95}. 

References \cite{Carmelo92a,Carmelo92b} include a study of
the charge and spin fluctuations in terms of the pseudoparticle 
excitations. However, that study did not use an operator
treatment of the problem. Therefore, it did not solve
the present and inverse problem of expressing the $c$ and
$s$ pseudoparticle excitation branches in terms of electronic
charge and spin operators. This latter problem requires
a careful analysis of the relevant Hilbert sub space where the
charge and spin fluctuations are contained. This study is
presented in the present paper in terms of the pseudoparticle
operator generators of the Hamiltonian eigenstates.

Our operator analysis allows the solution of an important 
open problem. This is the evaluation of the commutator between the 
electronic-density operators for the many-electron Hubbard chain. 
This quantity plays an important role in the physics of 
many-electron systems \cite{Anderson96}. We find the
exact expression of this commutator for the Hubbard chain
at finite values of the magnetic field and chemical
potential. Its spin-dependent factor is in general non 
diagonal and depends on the interaction, electronic
density, and spin density. This is the sign of the unconventional
$c$ and $s$ separation which in the presence of the magnetic
field replaces the usual charge and spin separation. 

The paper is organized as follows: In section II we
summarize some general features of the pseudoparticle
basis. The generators of the $c$ and $s$ low-energy and 
small-momentum excitations are expressed both in the 
pseudoparticle and electronic basis in Sec. III.
In section IV we explain the Luttinger-liquid character
of the Hubbard chain by our pseudoparticle approach:
we describe the low-energy physics in
terms of separate left and right pseudoparticle
conservation laws. The bosonization of the
$c$ and $s$ pseudoparticle excitations and the
evaluation of the commutator between the 
electronic-density operators are studied in
Sec. V. Finally, Sec. VI contains the concluding remarks.

\section{THE PSEUDOPARTICLE OPERATOR BASIS}

In order to study the effects of the magnetic field on the 
generators of the low-energy and small-momentum excitations of the 
1D integrable many-electron problem we consider in this paper the 
particular case of the Hubbard chain \cite{Lieb} in a magnetic field 
$H$ and chemical potential $\mu$ \cite{Carmelo94,Carmelo92,Carmelo92b} 

\begin{equation}
\hat{H} = -t\sum_{j,\sigma}
\left[c_{j\sigma}^{\dag }c_{j+1\sigma} +
c_{j+1\sigma}^{\dag }c_{j\sigma}\right] +
U\sum_{j} [c_{j\uparrow}^{\dag }c_{j\uparrow}-{1\over 2}]
[c_{j\downarrow}^{\dag }c_{j\downarrow}-{1\over 2}]
+ 2\mu\hat{\eta }_z + 2\mu_0 H\hat{S}_z \, ,
\end{equation}
where
\begin{equation}
\hat{\eta}_z = -{1\over 2}[N_a - \sum_{\sigma}\hat{N}_{\sigma }] \, ,
\hspace{2cm}
\hat{S}_z = -{1\over 2}\sum_{\sigma}\sigma\hat{N}_{\sigma } \, .
\end{equation}
Here $c_{j\sigma}^{\dagger}$ and $c_{j\sigma}$ are electron operators 
of spin projection $\sigma $ at site $j$, $\hat{N}_{\sigma }=\sum_{j} 
c_{j\sigma }^{\dagger }c_{j\sigma }$ and $t$, $U$, and $\mu _0$ are 
the transfer integral, the onsite Coulomb interaction, and the Bohr 
magneton, respectively. The Hamiltonian $(1)$ 
describes $N_{\uparrow}$ up-spin and $N_{\downarrow}$ down-spin
electrons in a lattice of $N_a$ sites ($N=N_{\uparrow}+
N_{\downarrow}$, $n=N/N_a$, $n_{\sigma }=N_{\sigma }/N_a$, 
and $m=n_{\uparrow}-n_{\downarrow}$). We consider in our study
electronic densities $0<n<1$ and spin densities 
$0<m<n$. This corresponds to a sector of parameter space with 
$U(1)\times U(1)$ symmetry \cite{Carmelo94,Carmelo92,Carmelo96}. 
In this case the low-energy physics is dominated by the 
lowest-weight states (LWS's) of both the eta-spin and 
spin algebras which refer to real rapidities 
\cite{Carmelo94,Carmelo92,Carmelo96}. We call these states 
``LWS's I'' to distinguish them from the LWS's associated with 
complex, non-real, rapidities, which we call ``LWS's II''. 
Importantly, in this $U(1)\times U(1)$ sector both the LWS's II 
and the non-LWS's have energy gaps relative to each canonical-ensemble
ground state \cite{Carmelo95} and do not contribute to the 
low-energy physics \cite{Carmelo94,Carmelo94a}. 

In the pseudoparticle basis \cite{Carmelo94,Carmelo94a} one can 
define a many-pseudoparticle perturbation theory in which the 
non-interacting pseudoparticle ground state of simple 
Slater-determinant form is the exact ground state of the 
many-electron problem. All LWS's I can be generated by acting 
on the vacuum $|V\rangle$ (zero-electron state) with pseudoparticle 
operators. This operator algebra involves two types 
of anti-commuting {\it pseudoparticle} creation (annihilation) operators 
$b^{\dag }_{q\alpha }$ ($b_{q\alpha }$) \cite{Carmelo94}. Here 
$\alpha$ refers to the two pseudoparticle colors $c$ and $s$ 
\cite{Carmelo94,Carmelo94a,Carmelo92b}. 

The discrete pseudomomentum values are $q_j={2\pi\over {N_a}}
I_j^{\alpha }$, where $I_j^{\alpha }$ are {\it consecutive} 
integers or half integers. There are $N_{\alpha }^*$ values of 
$I_j^{\alpha }$, {\it i.e.} $j=1,...,N_{\alpha }^*$. One LWS I 
is specified by the distribution of $N_{\alpha }$ occupied 
values, which we call $\alpha $ pseudoparticles, over the 
$N_{\alpha }^*$ available values. There are 
$N_{\alpha }^*-N_{\alpha }$ corresponding empty values, 
which we call $\alpha $ pseudoholes. In the case of the Hubbard 
chain, we have that 

\begin{equation}
N_c^*=N_a \, , \hspace{0.3cm} N_c=N \, , \hspace{0.3cm}
N_s^*=N_{\uparrow} \, , \hspace{0.3cm} N_s=N_{\downarrow} \, , 
\end{equation}
{\it i.e.}, the pseudoparticle numbers
are {\it good quantum numbers}. The boundary conditions fix the 
numbers $I_j^{\alpha }$: $I_j^c$ are integers (or half integers) 
for $N_{\downarrow}$ even (or odd), and $I_j^s$ are integers 
(or half integers) for $N_{\uparrow}$ odd (or even) 
\cite{Lieb,Carmelo94}. The ground state associated with a 
canonical ensemble of $(N_{\uparrow},N_{\downarrow})$ values 
[and $(N_c=N_{\uparrow}+N_{\downarrow},N_s=N_{\downarrow}$)]
has the form \cite{Carmelo94,Carmelo94a,Carmelo95} 

\begin{equation}
|0;N_{\uparrow},N_{\downarrow}\rangle = \prod_{\alpha=c,s}
[\prod_{q=q_{F\alpha }^{(-)}}^{q_{F\alpha }^{(+)}} 
b^{\dag }_{q\alpha }] 
|V\rangle \, .
\end{equation}
Here $b^{\dag }_{q\alpha }$ is the pseudoparticle operator 
of pseudomomentum $q$ and of color $\alpha =c,s$.
When $N_{\alpha }$ is odd (even) and $I_j^{\alpha }$ are 
integers (half integers) the pseudo-Fermi points are symmetric 
and given by $q_{F\alpha }^{(+)} = -q_{F\alpha }^{(-)}=
{\pi\over {N_a}}[N_{\alpha}-1]$. On the other hand, when 
$N_{\alpha }$ is odd (even) and $I_j^{\alpha }$ are half 
integers (integers) we have that $q_{F\alpha }^{(+)}=
{\pi\over {N_a}}N_{\alpha }$ and $-q_{F\alpha }^{(-)} =
{\pi\over {N_a}}[N_{\alpha }-2]$ or $q_{F\alpha }^{(+)}= 
{\pi\over {N_a}}[N_{\alpha }-2]$ and $-q_{F\alpha }^{(-)}= 
{\pi\over {N_a}}N_{\alpha }$

Except for terms of $1/N_a$
order, $q_{Fc }^{(\pm)}=\pm 2k_F=\pm \pi n$ and 
$q_{Fs }^{(\pm)}=\pm k_{F\downarrow}=\pm \pi n_{\downarrow}$.

Let $\cal {H}_I$ be the Hilbert space spanned by the LWS's 
I. In $\cal {H}_I$ and normal-ordered relative to the
ground state $(4)$, the Hubbard Hamiltonian $(1)$ has an infinite 
number of terms which correspond to increasing ``order'' of 
pseudoparticle scattering \cite{Carmelo94,Carmelo94a}. In the Appendix 
we present such Hamiltonian and the associate pseudoparticle 
parameters.

\section{THE SMALL-MOMENTUM AND LOW-ENERGY $c$ and $s$ GENERATORS}

Since the $c$ and $s$ pseudoparticles dominate the physics at
energy scales smaller than the gaps for the LWS's II and 
non-LWS's, it is clearly of interest to study the low-energy
and small-momentum generators in this region. The pseudoparticle 
operator algebra described in Refs. \cite{Carmelo94,Carmelo94b} is 
extracted directly from the BA solution. However, without the knowledge of 
the expressions of the Hamiltonian-eigenstate pseudoparticle 
generators in the electronic basis we cannot relate them to the 
usual charge and spin fluctuation operators. In this section we 
solve the problem for the case of the low-energy and small-momentum 
$c$ and $s$ generators which we express both in the pseudoparticle 
and electronic basis.

The usual charge and spin fluctuation operators read

\begin{equation}
{\hat{\rho}}_{\rho }(k) = {\hat{\rho}}_{\uparrow }(k) + 
{\hat{\rho}}_{\downarrow }(k) \, , \hspace{1.5cm}
{\hat{\rho}}_{\sigma_z }(k) = {\hat{\rho}}_{\uparrow }(k) - 
{\hat{\rho}}_{\downarrow }(k) \, ,
\end{equation}
respectively, where 

\begin{equation}
{\hat{\rho}}_{\sigma }(k)=
\sum_{k'}c^{\dag }_{k'+k\sigma}c_{k'\sigma}
\, ,
\end{equation}
and $c_{k\sigma}^{\dagger}$ and $c_{k\sigma}$ are electron operators 
of spin projection $\sigma $ and momentum $k$.

Our goal is expressing the $\alpha =c,s$ pseudoparticle
generators

\begin{equation}
{\hat{\rho}}_{\alpha }(k)=\sum_{q}b^{\dag }_{q+k\alpha}
b_{q\alpha} 
\, ,
\end{equation}
in the electronic basis. Since the $\alpha $ pseudoparticle
numbers are good quantum numbers of the many-electron
problem [see Eq. $(3)$], the problem is solved at
$k=0$. In this case the operators $(6)$ and $(7)$ just
provide the electronic and pseudoparticle number operators 
as follows

\begin{equation}
{\hat{N}}_{\sigma}={\hat{\rho}}_{\sigma }(0) 
\, , \hspace{1.5cm} 
{\hat{N}}_{\alpha}={\hat{\rho}}_{\alpha }(0) 
\, ,
\end{equation}
respectively. The relation between the pseudoparticle and 
electronic numbers follows directly from Eq. $(3)$ with
the result

\begin{equation}
{\hat{N}}_{\alpha} = \sum_{\sigma}G_{\alpha\sigma}
{\hat{N}}_{\sigma} \, , \hspace{1.5cm} 
{\hat{N}}_{\sigma} = \sum_{\sigma}
G^{-1}_{\sigma\alpha}{\hat{N}}_{\alpha}
\, ,
\end{equation}
where the ${\rm\bf G}$ and ${\rm\bf G}^{-1}$ 
electron - pseudoparticle and pseudoparticle - electron charge 
matrices, respectively, read 

\begin{equation}
{\rm\bf G} =
\left[\begin{array}{cc}
G_{c\uparrow} \hspace{0.5cm} & G_{c\downarrow} \\
G_{s\uparrow} \hspace{0.5cm} & G_{s\downarrow}
\end{array}
\right] =
\left[\begin{array}{cc}
1 \hspace{0.5cm} & 1 \\
0 \hspace{0.5cm} & 1
\end{array}
\right] = e^{\frac{1}{2} \sigma^+} \, ,
\end{equation}
and
\begin{equation}
{\rm\bf G}^{-1} =
\left[\begin{array}{cc}
1 \hspace{0.5cm} & -1 \\
0 \hspace{0.5cm} & 1
\end{array}
\right] = e^{- \frac{1}{2} \sigma^+}\, ,
\end{equation}
respectively, where $\sigma^+ = \sigma_x + i \sigma_y$ is a Pauli 
matrix. The simple form of these matrices follows from the 
conservation of the number of $\alpha$ pseudoparticles. However, 
that ${\hat{\rho}}_{c }(0)={\hat{\rho}}_{\rho}(0)$ and 
${\hat{\rho}}_{s }(0)={\hat{\rho}}_{\downarrow }(0)$ does 
{\it not} imply that such equalities hold true for $k>0$, as
we find below.

The pseudoparticle representation can be extended to
the whole Hilbert space. This requires the introduction
of new pseudoparticle modes which are absent at
low energy \cite{Carmelo96c}. However, at low-energy
there is a electron - pseudoparticle transformation
which refers only to the present $c$ and $s$ pseudoparticle
branches. An essential point is that this electron - pseudoparticle 
transformation {\it does not} mix left and right 
electronic operators, {\it i.e.}, $\iota =sgn (k)1=\pm 1$ 
electronic operators are made out of $\iota =sgn 
(q)1=\pm 1$ pseudoparticle operators only, $\iota $ defining 
the right ($\iota=1$) and left ($\iota=-1$) movers.

Measuring the electronic momentum $k$ and 
pseudomomentum $q$ from the $U=0$ Fermi points 
$k_{F\sigma}^{(\pm)}=\pm \pi n_{\sigma}$ and  pseudo-Fermi 
points $q_{F\alpha}^{(\pm)}$, respectively, adds the index 
$\iota$ to the electronic and pseudoparticle operators.
The new momentum $\tilde{k}$ and pseudomomentum $\tilde{q}$ are 
such that 

\begin{equation}
\tilde{k} =k-k_{F\sigma}^{(\pm)}
\, ,
\end{equation}
and

\begin{equation}
\tilde{q}=q-q_{F\alpha}^{(\pm)}
\, ,
\end{equation}
respectively, for $\iota=\pm 1$.
For instance, 

\begin{equation}
{\hat{\rho}}_{\rho\iota }(k) = {\hat{\rho}}_{\uparrow\iota }(k) + 
{\hat{\rho}}_{\downarrow\iota }(k) \, , \hspace{1cm} 
{\hat{\rho}}_{\sigma_z\iota }(k) = {\hat{\rho}}_{\uparrow\iota }(k) - 
{\hat{\rho}}_{\downarrow\iota }(k) \, ,
\end{equation}
where

\begin{equation}
{\hat{\rho}}_{\sigma\iota }(k)=\sum_{\tilde{k}}
c^{\dag }_{\tilde{k}+k\sigma\iota}c_{\tilde{k}\sigma\iota}
\, .
\end{equation}
Also

\begin{equation}
{\hat{\rho}}_{\alpha\iota }(k)=\sum_{\tilde{q}}b^{\dag 
}_{\tilde{q}+k\alpha\iota}b_{\tilde{q}\alpha\iota}
\, .
\end{equation}

In the Letter \cite{Carmelo94a} we have expressed the pseudoparticle
generators $(7)$ in terms of the electronic operators $(5)$
or $(6)$ by restricting our analysis to a reduced Hilbert
sub space. In order to justify the validity of that procedure
we consider here the whole low-energy Hilbert space. 
Let us express the one-pair electronic operators
$(6)$ at $k=\iota{2\pi\over {N_a}}$ in the pseudoparticle basis. 
Henceforth we denote the ground state $(4)$ by $|0\rangle $.
The excitation 

\begin{equation}
\hat{\rho}_{\sigma\iota}(k=\iota{2\pi\over 
{N_a}})|0\rangle \, , 
\end{equation}
is a superposition of both LWS's I, 
LWS's II, and non-LWS's. However, we are only interested
in the low-energy component of this excitation. Therefore,
we can omit here its LWS's II and non-LWS's components
which refer to energy scales larger than the gaps
of these states relative to the suitable ground state 
of form $(4)$.

In references \cite{Carmelo94,Carmelo95,Carmelo96} we have 
constructed the low-energy Hamiltonian eigenstates by acting 
suitable pseudoparticle generators onto the pseudoparticle
or pseudohole vacua. Following these studies, we can now 
describe the low-energy and momentum $k=\iota{2\pi\over {N_a}}$ 
Hilbert sub space where the excitation $(17)$ is contained. 
This space is spanned by all Hamiltonian eigenstates of the 
form

\begin{equation}
|\alpha\iota;N_{ph}^{c },N_{ph}^{s },l\rangle = 
{\hat{A}}_{N_{ph}^{c },N_{ph}^{s },l}|\alpha\iota\rangle \, ,
\end{equation}
where $N_{ph}^{\alpha }$ is the number of $\alpha $ 
pseudoparticle-pseudohole processes such that
${\sum_{\alpha}N_{ph}^{\alpha }\over N_a}\rightarrow 0$.
In the right-hand side (rhs) of Eq. $(18)$,

\begin{equation}
|\alpha\iota\rangle = \hat{\rho}_{\alpha\iota}(k=\iota 
{2\pi\over {N_a}})|0\rangle \, ,
\end{equation}
and

\begin{equation}
{\hat{A}}_{N_{ph}^{c },N_{ph}^{s },l} = \prod_{\alpha=c,s}
{\hat{L}}^{\alpha\iota}_{-N_{ph}^{\alpha\iota}}(l) \, , 
\end{equation}
where the operator ${\hat{L}}^{\alpha\iota}_{-N_{ph}^{\alpha\iota}}(l)$
is given in Eq. $(56)$ of Ref. \cite{Carmelo94b}.
It produces a number $N_{ph}^{\alpha ,\iota}$ of 
$\alpha ,\iota$ pseudoparticle-pseudohole processes. In the present
case ${\hat{A}}_{N_{ph}^{c },N_{ph}^{s },l}$ is a zero-momentum
operator and, therefore,

\begin{equation}
N_{ph}^{\alpha} = \sum_{\iota}N_{ph}^{\alpha ,\iota} = 
2N_{ph}^{\alpha ,\iota=1} = 
2N_{ph}^{\alpha ,\iota=-1} \, .
\end{equation}
In equations $(18)$ and $(20)$ $l$ is a quantum number which 
distinguishes different pseudoparticle-pseudohole distributions 
characterized by the same values for the numbers $(21)$.
It follows from Eq. $(21)$ that

\begin{equation}
\sum_{\alpha ,\iota}\iota
N_{ph}^{\alpha ,\iota} = 0 \, . 
\end{equation}
Since 

\begin{equation}
N_{ph}^{\alpha ,\iota}=1,2,3,.... \, ,
\end{equation}
equations $(21)-(22)$ imply that 

\begin{equation}
\sum_{\alpha }N_{ph}^{\alpha } = 
\sum_{\alpha ,\iota}
N_{ph}^{\alpha ,\iota} = 2,4,6,8.... \, , 
\end{equation}
is always an even positive integer.

The set of Hamiltonian eigenstates of form $(18)$ constitutes
a complete and orthonormal basis which spans at low-energy 
the Hilbert space where the excitation $(17)$ is defined. 
Therefore, at $k=\iota{2\pi\over {N_a}}$
it can be written as follows

\begin{equation}
\hat{\rho}_{\sigma\iota}(k)|0\rangle =
\sum_{\alpha,\iota,N_{ph}^{c },N_{ph}^{s },l}
\langle \alpha\iota;N_{ph}^{c },N_{ph}^{s },l|
\hat{\rho}_{\sigma\iota}(k)|0\rangle 
|\alpha\iota;N_{ph}^{c },N_{ph}^{s },l\rangle \, ,
\hspace{1cm} k=\iota{2\pi\over {N_a}} \, .
\end{equation}

The methods used in Refs. \cite{Carmelo92,Carmelo93,Carmelo96b}
allowed us deriving the matrix elements of the rhs
of Eq. $(25)$. In the present  
${\sum_{\alpha}N_{ph}^{\alpha }\over N_a}\rightarrow 0$ 
limit which corresponds to the limit of vanishing
excitation energy we find,

\begin{equation}
\langle \alpha\iota;N_{ph}^{c },N_{ph}^{s },l|
\hat{\rho}_{\sigma\iota}(k=\iota{2\pi\over 
{N_a}})|0\rangle = 0  \, ,
\end{equation}
for $N_{ph}^{c }>0$ or (and) $N_{ph}^{s }>0$ and

\begin{equation}
\langle\alpha\iota|\hat{\rho}_{\sigma\iota}(=\iota{2\pi\over 
{N_a}})|0\rangle =
\sum_{\alpha'}G^{-1}_{\sigma\alpha '}\xi^{1}_{\alpha\alpha'}
\, ,
\end{equation}
where the matrix entries $G^{-1}_{\sigma\alpha '}$ and 
$\xi^{1}_{\alpha\alpha'}$ are defined in Eqs. $(10)$ and
(A4), respectively. While the finite matrix elements
$(27)$ can be derived from a spectral analysis of the charge and 
spin response functions \cite{Carmelo92b}, the result
$(26)$ follows from the method used in Ref. \cite{Carmelo93}.
This is described for the particular case of 
one-electron matrix elements in Ref. \cite{Carmelo96b}.

We emphasize that the result $(26)$ corresponds to 
vanishing excitation energy whereas at low but finite energy 
there is overlap between the excitation $(17)$ and multi 
pseudoparticle-pseudohole Hamiltonian eigenstates. Equations
$(26)$ and $(27)$ tell us that in the limit of vanishing energy the 
excitation $(17)$ only couples to the one-pair 
pseudoparticle-pseudohole Hamiltonian eigenstates of the form 
$(19)$.

It is the vanishing-excitation-energy result $(26)-(27)$ which
justifies the validity of the procedure followed in 
Ref. \cite{Carmelo94a}: in order to find the expression of the 
$\alpha\iota $ pseudoparticle generators $(16)$ we can 
consider only the reduced Hilbert space spanned by the four 
Hamiltonian eigenstates of form $(19)$ and ignore all
the multi-pseudoparticle-pseudohole states $(18)$.
The states $(19)$ have momentum $k=\iota {2\pi\over 
{N_a}}$ relative to the ground state $(4)$ and have a single 
pseudohole at one of the pseudo-Fermi points. Obviously, these 
states are orthogonal to that ground state. In the reduced Hilbert 
space they constitute a complete orthonormal basis, so that 

\begin{equation}
\langle\alpha\iota|\alpha '\iota '\rangle =\delta_{\alpha
,\alpha '}\delta_{\iota ,\iota '} 
\, ,
\end{equation}
and

\begin{equation}
\sum_{\alpha ,\iota}|\alpha\iota\rangle\langle\alpha\iota|=
\openone
\, .
\end{equation}

Using equations $(25)-(29)$ we can write 
$\hat{\rho}_{\sigma\iota}(k)$ in the reduced
$k=\iota{2\pi\over {N_a}}$ Hilbert space with the result

\begin{equation}
\hat{\rho}_{\sigma\iota}(k)=\sum_{\alpha ,\iota}{\cal 
U}^{-1}_{\alpha\sigma}\hat{\rho}_{\alpha\iota}(k)
\, ,
\end{equation}
where we have introduced the pseudoparticle - electronic
matrix ${\rm\bf {\cal U}}^{-1}$ such that

\begin{equation}
{\cal U}^{-1}_{\alpha\sigma}=\langle\alpha\iota|
\hat{\rho}_{\sigma\iota}(k=\iota{2\pi\over {N_a}})|0\rangle
\, .
\end{equation}
Obviously, the result $(30)$ holds true only at 
$k=\iota{2\pi\over {N_a}}$ and vanishing excitation energy.

Let us now introduce the ``electronic'' states

\begin{equation}
|\sigma\iota\rangle=
\hat{\rho}_{\sigma\iota}(k=\iota{2\pi\over {N_a}})|0\rangle 
\, .
\end{equation}

Our task is finding the expression of the pseudoparticle
operators $\hat{\rho}_{\alpha\iota}(k)$ in the electronic basis. 
Fortunately, the $k=\iota{2\pi\over {N_a}}$ states $|\sigma\iota\rangle$ 
constitute a complete (but non-orthonormal) basis in the reduced 
Hilbert space spanned by the four states of form $(19)$. This
holds true only in the corresponding limit of
vanishing energy. At low but finite energy one finds out
that the one-pair pseudoparticle operators $(16)$ are both
of one-pair and multipair electronic character. However,
in the reduced Hilbert space we have that $\det 
{\rm\bf {\cal U}}^{-1}>0$ which implies that the set of
four electronic states of form $(32)$ constitutes a complete basis 
there. Therefore, we can invert the matrix ${\rm\bf {\cal U}}^{-1}$ 
and derive the following expression for the generator $(16)$,

\begin{equation}
\hat{\rho}_{\alpha\iota}(k) = 
\sum_{\sigma}{\cal U}_{\sigma\alpha} \hat{\rho}_{\sigma\iota}(k) \, ,
\hspace{1cm} k=\iota {2\pi\over {N_a}} \, ,
\end{equation}
where the electronic - pseudoparticle matrix 
${\rm\bf {\cal U}}$ is such that 

\begin{equation}
{\cal U}_{\sigma\alpha}=
\sum_{\alpha'}G_{\alpha'\sigma}\xi^{0}_{\alpha\alpha'} 
\, ,
\end{equation}
with the entries $G_{\alpha'\sigma}$ and $\xi^{0}_{\alpha\alpha'}$
defined in Eqs. $(11)$ and (A4), respectively. We can alternatively 
express the pseudoparticle generators $(16)$ in terms of the charge 
and spin fluctuation operators $(14)$ with the result

\begin{equation}
\hat{\rho}_{\alpha\iota}(k) = 
{\cal U}_{\rho\alpha} \hat{\rho}_{\rho\iota}(k) 
+ {\cal U}_{\sigma_z\alpha} \hat{\rho}_{\sigma_z\iota}(k)
\, ,
\hspace{1cm} k=\iota {2\pi\over {N_a}} \, ,
\end{equation}
where

\begin{equation}
{\cal U}_{\rho\alpha} = {1\over 2}\sum_{\alpha',\sigma =\pm 1}
G_{\alpha'\sigma}\xi^{0}_{\alpha\alpha'} 
\, ,  \hspace{1cm}
{\cal U}_{\sigma_z\alpha} = {1\over 2}\sum_{\alpha',\sigma =\pm 1}
\sigma G_{\alpha'\sigma}\xi^{0}_{\alpha\alpha'} 
\, .
\end{equation}

Equations $(35)$ and $(36)$ reveal that the generators of the 
$\alpha$ pseudoparticle excitations are an interaction-dependent 
mixture of charge and spin fluctuation operators. This 
holds true at finite values of the magnetic field and
away of half filling where {\it all} the 
$2\times 2$ matrix elements ${\cal U}_{\rho\alpha}$ 
and ${\cal U}_{\sigma_z\alpha}$ are non-vanishing and 
interaction dependent. This shows the exotic character of the 
$c$ - $s$ low-energy separation. 

However, from the usual $m=0$ and (or) $n=1$ pictures 
\cite{Castro,Haldane,Schulz}, we expect $c$ and
$s$ to become charge and spin, respectively, in the limit
$m\rightarrow 0$, and $c$ to become charge in the limit 
$n\rightarrow 1$. This is confirmed by the data presented in
the Table where we show some limiting forms of the pseudoparticle
generators $(35)$. We thus conclude that while at zero 
magnetic field there is the well known charge - spin
separation of the low-energy and small-momentum excitations
of the Hubbard chain, the presence of the magnetic field 
changes the pre-existent charge and spin modes into new $c$ 
and $s$ separate excitation modes, respectively, which couple 
both to charge and spin. Our expression $(35)$ fully characterizes
the generators of these excitations in terms of the usual
charge and spin fluctuation operators.

\section{LOW-ENERGY PSEUDOPARTICLE LEFT AND RIGHT 
CONSERVATION LAWS}

Following the results of the Letter \cite{Carmelo94a},
the occurence at low-energy of separate right and left
conservation laws \cite{Castro} refers in BA solvable
problems to the pseudoparticle scattering.

In this section we show that the Luttinger-liquid character
of 1D interacting models solvable by BA follows from the
properties of the Hamiltonian in the pseudoparticle
operator basis. The pseudoparticle operator algebra is used 
to construct a general Luttinger-liquid theory. The non-interacting
reference ground state of this theory is the Slatter determinant
of pseudoparticle levels given in Eq. $(4)$. This is the
ground state of the many-electron problem. Moreover,
the construction of the critical-point Hamiltonian
proceeds by linearizing the pseudoparticle bands 
instead of the electronic bands.

Let us show that the pseudoparticle Hamiltonian
(A1)-(A2) is the correct starting point to study the low-energy
physics of the many-Hamiltonian $(1)$ in terms
of right and left conservation laws \cite{Castro,Carmelo94a}.
It also follows that in the presence of the magnetic field
the low-energy excitations can be bosonized. However,
this bosonization corresponds to the $c$ and $s$ excitation
modes studied in the previous section and refers to the 
pseudoparticle basis. In contrast to the zero-field case (where 
$c$ and $s$ are charge and spin, respectively -- see the Table), 
the present bosonization scheme cannot be easily described in 
terms of charge and spin fluctuations. This follows from the exotic
form of the generators $(16)$ in terms of such fluctuations,
Eq. $(35)$.

While the two-electron amplitudes of scattering diverge 
at the Fermi points, the two-pseudoparticle amplitudes
of forward scattering and corresponding $f$ functions are
finite and determine the low-energy parameters 
\cite{Carmelo94,Carmelo94a,Carmelo92,Carmelo96b}. Therefore,
the quantum problem $(1)$ is non perturbative and perturbative
in the electronic and pseudoparticle basis, respectively.
This perturbative character of the pseudoparticle basis 
implies that the Hamiltonian (A1)-(A3) can be used and 
is the more suitable as starting point for the construction of 
a critical-point Hamiltonian. 

This proceeds by linearizing the pseudoparticle bands 
$\epsilon_{\alpha}(q)$ around the pseudo-Fermi points and 
replacing the full $f$ functions by 

\begin{equation}
f_{\alpha\alpha'}^{1}=
f_{\alpha\alpha'}(q_{F\alpha}^{(\pm)},
q_{F\alpha'}^{(\pm)})
\, ,
\end{equation}
or 

\begin{equation}
f_{\alpha\alpha'}^{-1}= 
f_{\alpha\alpha'}(q_{F\alpha}^{(\pm)},q_{F\alpha'}^{(\mp)})
\, .
\end{equation}
The expressions of $f_{\alpha\alpha'}^{\pm 1}$ are simple combinations
of the pseudoparticle Landau parameters defined in Ref.
\cite{Carmelo92} and involve the velocities $v_{\alpha}$ and 
parameters $\xi_{\alpha\alpha '}^j$ of Eq. (A4) only. It
is the form of this critical point Hamiltonian which implies
the Luttinger-liquid character of the low-energy spectrum 
provided by BA and conformal-field theory \cite{Frahm}.
The critical-point Hamiltonian can be written as

\begin{equation}
:\hat{{\cal H}}:={\hat{{\cal H}}}_0 + {\hat{{\cal H}}}_2
+ {\hat{{\cal H}}}_4 \, ,
\end{equation}
where

\begin{equation}
{\hat{{\cal H}}}_0 = 
\sum_{\alpha ,\iota ,\tilde{q}}
\iota v_{\alpha}\tilde{q}  :\hat{N}_{\alpha,\iota}(\tilde{q}):
\, ,
\end{equation}
and

\begin{equation}
{\hat{{\cal H}}}_2 + {\hat{{\cal H}}}_4 =   
{2\over {N_a}}\sum_{\alpha,\alpha',\iota,\iota',k} 
[g_2^{\alpha\alpha '}(k)\delta_{\iota ,-\iota '} + 
g_4^{\alpha\alpha '}(k)\delta_{\iota ,\iota '}]
:{\hat{\rho}}_{\alpha ,\iota}(k):
:{\hat{\rho}}_{\alpha ',\iota'}(-k):
\, .
\end{equation}
Here

\begin{equation}
{\hat{N}}_{\alpha\iota}(\tilde{q})=
b^{\dag }_{\tilde{q}\alpha\iota}b_{\tilde{q}\alpha\iota}
\, ,
\end{equation}
and the couplings read 

\begin{equation}
g_2^{\alpha\alpha '}(k)=
f_{\alpha\alpha'}^{-1}\delta_{k,0}
\, ,
\end{equation}
and 

\begin{equation}
g_4^{\alpha\alpha '}(k)=
f_{\alpha\alpha'}^{1}\delta_{k,0}
\, .
\end{equation}
It follows that in the pseudoparticle basis the suitable 
critical-point Hamiltonian is a g-ology model of the type 
studied in Refs. \cite{Solyom,Castro} with exotic $k=0$ 
forward-scattering couplings, $\sigma$ replaced by $\alpha$, 
and the electronic operators by pseudoparticle operators.

This allows the ground-state distribution, $\langle 
\hat{N}_{\alpha}(q)\rangle $, to be equal to 1 inside and 0 
outside the pseudo-Fermi surface \cite{Carmelo94,Carmelo92,Carmelo95},  
as confirmed by Eq. $(4)$. The absence of the $g_1$ and $g_3$ 
terms implies that the $\alpha,\iota$ pseudoparticle number 
operators, 

\begin{equation}
\hat{N}_{\alpha,\iota}=\sum_{\tilde{q}}
\hat{N}_{\alpha,\iota}(\tilde{q})
\, ,
\end{equation}
are good quantum numbers, {\it i.e.} there are {\it 
separate} right and left conservation laws. This is a 
generalization of the results of Ref. \cite{Castro} with 
the Fermi points replaced by the pseudo-Fermi points.
In the case of single-component models 
\cite{Haldane,Anto}, we can omit the index $\alpha$, so 
that the BA critical Hamiltonian can be rewritten as 

\begin{equation}
:\hat{{\cal H}}:={\hat{{\cal H}}}_0+\hat{V}
\, ,
\end{equation}
with

\begin{equation}
\hat{V}={\pi\over {N_a}}\sum_{\iota ,\iota ',k} 
[V_1(k)\delta_{\iota ,\iota '}+V_2(k)\delta_{\iota ,-\iota 
'}]:{\hat{\rho}}_{\iota}(k)::{\hat{\rho}}_{\iota'}(-k):
\, .
\end{equation}
Here 

\begin{equation}
{\hat{\rho}}_{\iota}(k)=\sum_{\tilde{q}}b^{\dag 
}_{\tilde{q}+k\iota}b_{\tilde{q}\iota}
\, ,
\end{equation}

\begin{equation}
V_1(k)={f^{1}\over {2\pi}}\delta_{k,0}
\, ,
\end{equation}
and

\begin{equation}
V_2(k)={f^{-1}\over {2\pi}}\delta_{k,0}
\, .
\end{equation}
Therefore, in {\it single-component} integrable systems, 
$:\hat{{\cal H}}:$ is a Luttinger model with exotic $k=0$ 
forward-scattering potentials. This justifies the 
Luttinger-liquid character of integrable models by BA. 
However, we have confirmed here that such character corresponds
in the integrable many-electron problems to the pseudoparticle
basis.

The separate right and left conservation laws provide
the Luttinger-liquid parameters through equations
of motions \cite{Castro,Carmelo94a}. Let $\vartheta $ denote charge 
($\vartheta =\rho$) or spin ($\vartheta =\sigma_z$). 
Then ${\hat{N}}_{\vartheta\iota}$ can be written as 

\begin{equation}
{\hat{N}}_{\vartheta\iota}=\sum_{\alpha}
k_{\vartheta\alpha}{\hat{N}}_{\alpha\iota}
\, ,
\end{equation}
where the integers $k_{\vartheta\alpha}$ are $k_{\rho 
c}=k_{\sigma_{z} c}=1$, $k_{\rho s}=0$, and $k_{\sigma_{z} s}=-2$. 
$k_{\uparrow s}=-1$. 

Let us consider the charge or spin operator $(14)$ which
in the present notation are referred as ${\hat{\rho}}_{\vartheta\iota 
}(k)$. It is useful to consider the combinations 

\begin{equation}
{\hat{\rho}}_{\vartheta }^{(\pm)}(k)=
{\hat{\rho}}_{\vartheta 1}(k) \pm {\hat{\rho}}_{\vartheta -1}(k)
\, ,
\end{equation}
(note that ${\hat{\rho}}_{\vartheta }^{(+)}(k)=
{\hat{\rho}}_{\vartheta }(k)$). Since the commutator

\begin{equation}
[{\hat{\rho}}_{\vartheta }^{(\pm)}(k,t),
{\hat{{\cal H}}}_2+{\hat{{\cal H}}}_4]=0 
\, ,
\end{equation}
for $k>0$ and

\begin{equation}
[{\hat{\rho}}_{\vartheta }^{(\pm)}(k,t),:\hat{{\cal H}}:]
\, ,
\end{equation}
is proportional to $k$ at $k=0$, the interesting quantity associated 
with the equation of motion for 
${\hat{\rho}}_{\vartheta }^{(\pm)}(k,t)$ \cite{Castro,Carmelo94a} is 
the following ratio

\begin{equation}
{i\partial_t {\hat{\rho}}_{\vartheta }^{(\pm)}(k,t)\over 
k}|_{k=0} = {[{\hat{\rho}}_{\vartheta }^{(\pm)}(k,t),
:\hat{{\cal H}}:]\over k}|_{k=0} ={\cal 
V}_{\vartheta}^{(\mp)}{\hat{\rho}}_{\vartheta 
}^{(\mp)}(0,t) 
\end{equation}
where 

\begin{equation}
{\cal V}_{\vartheta}^{(-)} =
\sum_{\alpha ,\alpha'}k_{\vartheta\alpha}k_{\vartheta\alpha'}
[v_{\alpha}\delta_{\alpha ,\alpha '} + {f_{\alpha\alpha '}^{1}-
f_{\alpha\alpha '}^{-1}
\over {2\pi}}] = \sum_{\alpha}v_{\alpha}
[\sum_{\alpha'}k_{\vartheta\alpha '}\xi_{\alpha\alpha '}^1]^2
\, ,
\end{equation}
and 

\begin{equation}
{\cal V}_{\vartheta}^{(+)} = {1\over
{{\sum_{\alpha ,\alpha'}{k_{\vartheta\alpha}k_{\vartheta\alpha'}
\over {v_{\alpha}v{\alpha '}}}
[v_{\alpha}\delta_{\alpha ,\alpha '} - {A_{\alpha\alpha '}^{1}+
A_{\alpha\alpha '}^{-1}
\over {2\pi}}}]}} = {1\over {\{\sum_{\alpha}{1\over {v_{\alpha}}}
[\sum_{\alpha'}k_{\vartheta\alpha '}\xi_{\alpha\alpha '}^1]^2\}}}
\, .
\end{equation}
Here $A_{\alpha\alpha'}^{1}=A_{\alpha\alpha'
}(q_{F\alpha}^{(\pm)}, q_{F\alpha'}^{(\pm)})$ and 
$A_{\alpha\alpha'}^{-1}= A_{\alpha\alpha'}(q_{F\alpha}^{(\pm)},
q_{F\alpha'}^{(\mp)})$, where $A_{\alpha\alpha'}(q,q')$ are 
the scattering amplitudes given by Eqs. $(83)-(85)$ of 
Ref. \cite{Carmelo92b}. The velocities 
${\cal V}_{\vartheta}^{(+)}$ and ${\cal V}_{\vartheta}^{(-)}$ determine 
the $\vartheta$ susceptibility, $K^{\vartheta }=1/[\pi{\cal 
V}_{\vartheta}^{(+)}]$, and the coherent part of the 
$\vartheta $ conductivity spectrum, ${\cal V}_{\vartheta}^{(-)}
\delta (\omega )$, respectively \cite{Castro,Carmelo94a}. This 
agrees with the studies of Refs. \cite{Carmelo92,Carmelo92b}. 

For single-component systems \cite{Anto} there is only one 
choice for $\vartheta$ and ${\cal V}^{(-)}=v[\xi^1]^2$ 
and ${\cal V}^{(+)}=v[\xi^0]^2=v/[\xi^1]^2$, in agreement
with Ref. \cite{Haldane}. The ${\cal V}_{\vartheta}^{(\pm)}$ 
are the expressions for the charge and spin ``velocities''
of integrable {\it multicomponent} Luttinger liquids. 

Equation $(54)$ involves the commutator of the pseudoparticle-Hamiltonian 
$:\hat{{\cal H}}:$, Eq. $(39)$, with an electronic operator and, therefore, 
the velocities ${\cal V}_{\vartheta}^{(\pm)}$ do not have the 
same simple form as for the g-ology model of Ref. \cite{Castro}. 
Importantly, except for single-component integrable models, 
${\cal V}_{\vartheta}^{(+)}$ {\it does not} equal the 
expression of ${\cal V}_{\vartheta}^{(-)}$ 
with $f_{\alpha\alpha '}^{1}-f_{\alpha\alpha '}^{-1}$
replaced by $f_{\alpha\alpha '}^{1}+f_{\alpha\alpha '}^{-1}$.

\section{PSEUDOPARTICLE BOSONIZATION}

The bosonization of the critical-point Hamiltonian $(39)$
is straightforward and refers to the non-interacting pseudoparticle 
ground state $(4)$. We find that 

\begin{equation}
[\hat{\rho}_{\alpha\iota}(k),
\hat{\rho}_{\alpha'\iota'}(-k')]=\delta_{\alpha,\alpha'}
\delta_{\iota ,\iota'}\delta_{k,k'}(\iota
k{N_a\over {2\pi}}) 
\, ,
\end{equation}
and the $\alpha$ bosonic operators are given by

\begin{equation}
a^{\dagger}_{k\alpha} = \sqrt{{2\pi\over 
{N_a|k |}}} \sum_{\iota} \theta (\iota k) 
\hat{\rho}_{\alpha\iota}(k)  \, ,
\end{equation}
for $k>0$. 

This bosonization reproduces the results of conformal-field 
theory \cite{Frahm,Carmelo94b}: the bosons $(59)$ refer to the tower 
excitations of Ref. \cite{Carmelo94b}, whereas the HWS's 
\cite{Carmelo94b} of the Virasoro Algebras \cite{Frahm} are the 
current and ``charge'' excitations \cite{Haldane}. The low-energy 
separation refers to the colors $\alpha $ studied in Sec. III for 
all parameter space. Therefore, in the presence of the
magnetic field and chemical potential our boson modes do
not correspond to charge and spin but to the $c$ and $s$
excitations of exotic generators given by Eq. $(35)$.
It follows that in the electronic basis and at $k=\iota 
{2\pi\over {N_a}}$ the operators $(59)$ read 

\begin{equation}
a^{\dagger}_{k\alpha} =
\sqrt{{2\pi\over {N_a|k |}}} \sum_{\iota} \theta 
(\iota k)\left[{\cal U}_{\rho\alpha}\hat{\rho}_{\rho\iota}(k)
+{\cal U}_{\sigma_z\alpha}\hat{\rho}_{\sigma_z\iota}(k)\right]
\, ,
\end{equation}
with ${\cal U}_{\rho\alpha}$ and ${\cal U}_{\sigma_z\alpha}$
defined in Eq. $(36)$.

The commutator $(58)$ involves the ``pseudoparticle-density''
operators. On the other hand, the evaluation of the commutation 
relations between the electronic-density operators $(15)$ 
is a problem of high physical interest. The corresponding 
commutator has an important role in the physics of a many-electron 
quantum liquid \cite{Anderson96}. The results we have obtained in 
previous sections allow the exact solution of this problem for
the Hubbard chain in a magnetic field and chemical
potential. In order to calculate these commutation relations  
between the electronic-density operators $(15)$ we use Eqs. $(30)$ 
and $(58)$ to find

\begin{equation}
[\hat{\rho}_{\sigma \iota}(k),
\hat{\rho}_{\sigma'\iota'}(-k')]=\Delta_{\sigma,\sigma'}
\delta_{\iota ,\iota'}\delta_{k,k'}(\iota
k{N_a\over {2\pi}}) \, .
\label{electroncom}
\end{equation}
This commutator is related to the electronic finite
on-shell forward scattering phase shift referred in
Ref. \cite{Anderson96}. However, the matrix 

\begin{equation}
\Delta = 
\left[\begin{array}{cc}
\Delta_{\uparrow,\uparrow}
\hspace{0.5cm} & 
\Delta_{\uparrow,\downarrow}\\
\Delta_{\downarrow,\uparrow}
\hspace{0.5cm} & 
\Delta_{\downarrow,\downarrow}
\end{array}
\right]
\label{def} \, ,
\end{equation}
has not been evaluated for the Hubbard chain. We find 
it is non-diagonal and interaction dependent. Its entries 
read

\begin{equation}
\Delta_{\sigma,\sigma'} = \sum_{\alpha} {\cal U}^{-1}_{\alpha\sigma} {\cal
U}^{-1}_{\alpha \sigma'} \, .
\end{equation}
Referring directly to the above electronic phase shifts,
it is natural that this matrix is also determined by the pseudoparticle
forward-scattering interactions. Indeed, it can be 
expressed in terms of the simple combinations of 
two-pseudoparticle forward-scattering phase shifts, parameters (A4). 
The use of expression $(27)$ which defines the quantity $(31)$ 
leads to

\begin{equation}
\Delta = 
\left[\begin{array}{cc}
\left(\xi^{1}_{cc}-\xi^{1}_{cs}\right)^2 + 
\left(\xi^{1}_{sc}-\xi^{1}_{ss}\right)^2
\hspace{0.5cm} & 
\left(\xi^{1}_{cc}-\xi^{1}_{cs}\right)\xi^{1}_{cs} + 
\left(\xi^{1}_{sc}-\xi^{1}_{ss}\right)\xi^{1}_{ss}\\
\left(\xi^{1}_{cc}-\xi^{1}_{cs}\right)\xi^{1}_{cs} + 
\left(\xi^{1}_{sc}-\xi^{1}_{ss}\right)\xi^{1}_{ss} 
\hspace{0.5cm} & 
\left(\xi^{1}_{cs}\right)^2 + 
\left(\xi^{1}_{ss}\right)^2
\end{array}
\right] \, ,
\label{ex}
\end{equation}
where the $\xi^{1}_{\alpha\alpha '}$'s are the entries of
the matrix

\begin{equation}
\xi^{1} = 
\left[\begin{array}{cc}
\xi^{1}_{cc}
\hspace{0.5cm} & 
\xi^{1}_{cs}\\
\xi^{1}_{sc}
\hspace{0.5cm} & 
\xi^{1}_{ss}
\end{array}
\right] \, ,
\label{outra}
\end{equation}
which is such that

\begin{equation}
\det[\Delta] = \left(\det[\xi^{1}]\right)^2 \, .
\end{equation}
The entries $\xi^{1}_{\alpha\alpha '}$ are the above
simple combinations of two-pseudoparticle zero-momentum forward-scattering
phase shifts defined by Eq. (A4). Therefore, the electronic matrix 
$(64)$ is fully determined by the corresponding two-pseudoparticle 
forward-scattering collisions. This confirms that there is a direct 
relation between the electronic and pseudoparticle forward 
scattering \cite{Carmelo96b}.

We have just evaluated exactly the quantity 
$\Delta_{\sigma,\sigma'}$ of the commutator $(61)$. Our expression 
$(64)$ refers to electronic densities $0<n<1$ and spin densities 
$0<m<n$. In this case the Hamiltonian symmetry is $U(1)\otimes U(1)$.
In order to prove the expected $SU(2)$ invariance of the
matrix $\Delta $ at zero magnetic field \cite{Anderson96} we
should consider several directions in spin space. This study
will be presented elsewhere. However, effects of the $SU(2)$ spin 
symmetry show up in the limit $H\rightarrow 0$ of our present
expression.

The interaction, density, and spin-density dependence of $(64)$ 
is of important physical meaning. It describes the dependence
on these parameters of the finite electronic on-shell
forward-scattering phase shift \cite{Anderson96}.
In order to study the matrix $(64)$ in 
different limits we present in the Appendix the corresponding 
limiting values for the matrix $(65)$. Based on the expressions of the
Appendix we find that when $U\rightarrow 0$ for $\mu >W /2$ 
(and thus $n<1$) and $H>0$ (and thus $m>0$) [$W$ is 
the Mott-Hubbard gap] $(64)$ reduces to

\begin{equation}
\Delta_{\sigma,\sigma'} = \delta_{\sigma ,\sigma'} \, .
\end{equation}

Let us now consider the zero-magnetic-field and half-filling limits 
of expression $(64)$. This study reveals properties which are
a direct manifestation of the spin $SU(2)$ symmetry at $H=0$.
For instance, that symmetry implies that in the limit $H\rightarrow 0$
and $\mu >W /2$ the following form for the entries of $(64)$

\begin{eqnarray}
\Delta_{\sigma,\sigma'} & = & \delta_{\sigma ,\sigma'} 
- {\left[1 - \det[\Delta]\right]\over 2}\nonumber \\
& = & \delta_{\sigma ,\sigma'} 
- {1\over 2}\left[1 -{\left(\xi_0\right)^2\over 2}\right]  \, ,
\end{eqnarray}
where the parameter $\xi_0$ was studied in Ref. \cite{Carmelo92}.
We recall that for densities $0<n<1$ it changes from $\sqrt{2}$
for $U\rightarrow 0$, to $1$ as $U\rightarrow\infty$. Therefore,
$\det[\Delta]={\left(\xi_0\right)^2\over 2}$ changes from $1$
for $U\rightarrow 0$, to ${1\over 2}$ as $U\rightarrow\infty$. 
The fact that in the present limit the entries $(68)$ decouple in 
the spin-dependent $\delta_{\sigma ,\sigma'}$ simple term
and in a $\sigma ,\sigma'$ independent term follows from the
zero-field $SU(2)$ spin symmetry. Note also that the latter
term arises because of the removal by the electron-electron 
interaction of density spectral weight from low energy. This
is associated with values $\det[\Delta]<1$ for $U>0$.
Obviously, in the limit $U\rightarrow 0$ expression $(68)$ reads

\begin{equation}
\Delta_{\sigma,\sigma'} = \delta_{\sigma ,\sigma'} \, .
\end{equation}
Note that the $H=0$ $SU(2)$ spin symmetry has other effects on 
the $H\rightarrow 0$ limit. For instance, the $H\rightarrow 0$, 
$U\rightarrow 0$ and $U\rightarrow 0$, $H\rightarrow 0$ limits
of $\xi^{1}_{\alpha\alpha '}$ (see the Appendix)
do not commute. However, we emphasize that this does not show 
up in the quantity $(64)$ which has the same value in both 
these limits.

When $H\rightarrow 0$ and $U\rightarrow\infty$ for $\mu >W /2$ 
we find

\begin{equation}
\Delta_{\sigma,\sigma'} = \delta_{\sigma ,\sigma'} 
- 1/4 \, .
\end{equation}

At finite values of the magnetic field the entries of $(64)$
have not the simple form $(68)$. For instance,
in the $H\rightarrow H_c$ fully polarized limit (the critical field 
for onset of full polarized ferromagnetism, $H_c$, is given by 
the rhs of Eq. $(2)$ of Ref. \cite{Carmelo92b}) the result is

\begin{equation}
\Delta = 
\left[\begin{array}{cc}
1 + (\eta_0 -1)^2 
\hspace{0.5cm} & 
\eta_0 -1\\
\eta_0 -1
\hspace{0.5cm} & 
1
\end{array}
\right] \, ,
\end{equation}
where $\eta_0={2\over {\pi}}\tan^{-1}({\sin (\pi n)\over u})$. 

When $\mu\rightarrow W /2$ (half filling) the matrix $(64)$ 
reads

\begin{equation}
\Delta = 
\left[\begin{array}{cc}
\left(1-\xi^{1}_{cs}\right)^2 + 
\left(\xi^{1}_{ss}\right)^2
\hspace{0.5cm} & 
\left(1-\xi^{1}_{cs}\right)\xi^{1}_{cs} 
-\left(\xi^{1}_{ss}\right)^2\\
\left(1-\xi^{1}_{cs}\right)\xi^{1}_{cs} 
-\left(\xi^{1}_{ss}\right)^2
\hspace{0.5cm} & 
\left(\xi^{1}_{cs}\right)^2 + 
\left(\xi^{1}_{ss}\right)^2
\end{array}
\right] \, .
\end{equation}
If we take both the $\mu\rightarrow W /2$ and $H\rightarrow 0$ 
limits we find

\begin{equation}
\Delta_{\sigma,\sigma'} = \delta_{\sigma ,\sigma'} 
- 1/4 \, .
\end{equation}
If we consider the $\mu\rightarrow W /2$
and $H\rightarrow H_ c$ limits the result is

\begin{equation}
\Delta = 
\left[\begin{array}{cc}
2 
\hspace{0.5cm} & 
- 1\\
-1
\hspace{0.5cm} & 
1
\end{array}
\right] \, .
\end{equation}

Finally, in the limit $\mu\rightarrow W /2+4t$
(and thus $n\rightarrow 0$, which implies zero-electron density) 
we find when $H>0$ 

\begin{equation}
\Delta = 
\left[\begin{array}{cc}
2 
\hspace{0.5cm} & 
- 1\\
-1
\hspace{0.5cm} & 
1
\end{array}
\right] \, ,
\end{equation}
whereas if we take first the limit $H\rightarrow 0$ and then
$\mu\rightarrow W /2+4t$ we arrive to

\begin{equation}
\Delta_{\sigma,\sigma'} = \delta_{\sigma ,\sigma'} 
- 1/4 \, .
\end{equation}
In this case the $H=0$ spin $SU(2)$ symmetry has a direct
effect on the limiting values of $(64)$ which are
different in the limits $H\rightarrow 0$, $\mu\rightarrow W /2+4t$ and
$\mu\rightarrow W /2+4t$, $H\rightarrow 0$.

\section{CONCLUDING REMARKS}

In this paper we have expressed the generators of the $c$ and
$s$ low-energy and small momentum excitation modes of the
Hubbard chain in the electronic basis. This has revealed that
such eigenstate excitation branches are an interaction, density,
and magnetization dependent mixture of the charge and spin
fluctuation operators. As it is illustrated by the data of
the Table, in the limit of zero magnetic field the $c$ and
$s$ generators become the charge and spin fluctuation
operators, respectively, and in the limit of half filling
(zero chemical potential) the $c$ generator becomes
the charge fluctuation operator. Following that Table,
when we approach the limit $U\rightarrow 0$ at finite
values of the magnetic field [colun (a) in the Table]
and at zero magnetic field [colun (b) in the Table]
the $c$ and $s$ generators become different operators.
This effect of the magnetic field follows from the
removal of the spin $SU(2)$ symmetry and was already
detected in Ref. \cite{Carmelo92b}. Note also that
in the limits of zero electronic density and zero
magnetization the $s$ generator becomes the down-spin 
generator $(15)$, whereas in the former limit the $c$ 
generator becomes the charge fluctuation operator $(14)$. 

Our study has also revealed that the Landau-liquid character
of the pseudoparticle representation does not contradict
the Luttinger-liquid character of the BA energy spectrum.
(Also the conformal character of the energy spectrum
of electronic BA solvable models \cite{Frahm} follows
from the form of the quantum problem Hamiltonian in
the pseudoparticle basis \cite{Carmelo94b,Carmelo92}.)
The Landau-liquid character refers to the description
of the low-energy physics in terms of pseudoparticles with 
only zero-momentum and zero-energy forward scattering
interactions \cite{Carmelo94,Carmelo92}. We have shown that
the construction of the correct critical-point Hamiltonian
follows from the Hamiltonian written in the Landau-liquid
pseudoparticle representation. The exotic zero-momentum
forward-scattering character of the $g$ couplings $(43)$ and 
$(44)$ and potentials $(49)$ and $(50)$ follows directly from
the above Landau character of the quantum problem.
These Luttinger-liquid parameters are fully determined
by the values of the general Landau-liquid 
forward-scattering $f$ functions at the pseudo-Fermi points.

The obtained Luttinger-liquid theory and associate
bosonization refers to the pseudoparticle non-interacting
ground state $(4)$. However, we emphasize that in the
electronic basis this is the exact many-electron ground
state. Importantly, our study has allowed the evaluation
of the exact expression of the commutator between
electronic-density operators for densities $n<1$ and
spin densities $m>0$. We have found that this important
quantity \cite{Anderson96} is interaction dependent.
In the present work we have not considered different 
directions in spin space at zero magnetic field. 
However, the study of that commutator in the limit
$H\rightarrow 0$ has revealed interesting effects of
the zero-field $SU(2)$ spin symmetry.

\nonum
\section{ACKNOWLEDGMENTS}

We thank D. K. Campbell and N. M. R. Peres for stimulating 
discussions. This research was supported by the ISI Foundation 
(Torino) under the EU Contract No. ERBCHRX - CT920020.

\vfill
\eject
\appendix{THE MANY-PSEUDOPARTICLE HAMILTONIAN}

In the low-energy Hilbert sub space $\cal {H}_I$ introduced in 
Sec. II and in normal-ordered relative to the ground state of
form $(4)$, the Hubbard Hamiltonian $(1)$ reads 
\cite{Carmelo94,Carmelo92}

\begin{equation}
:\hat{H}:=\sum_{i=1}^{\infty} \hat{H}^{(i)} \, ,
\end{equation}
where the first two terms are

\begin{equation}
\hat{H}^{(1)}=\sum_{q,\alpha}\epsilon_{\alpha}(q)
:\hat{N}_{\alpha}(q):
\, ,
\end{equation}
and 

\begin{equation}
\hat{H}^{(2)}={1\over 
{N_a}}\sum_{q,\alpha} \sum_{q',\alpha'}{1\over 2}
f_{\alpha\alpha'}(q,q'):\hat{N}_{\alpha}(q):
:\hat{N}_{\alpha'}(q'):
\, .
\end{equation}

The expressions for the pseudoparticle bands, $\epsilon_{\alpha}(q)$, 
and of the ``Landau'' $f$ functions, $f_{\alpha\alpha'}(q,q')$, are 
given in Refs. \cite{Carmelo91b} and \cite{Carmelo92}, respectively. 
The latter involve the velocities 
$v_{\alpha}(q) = {d\epsilon_{\alpha}(q)\over {dq}}$
and the two-pseudoparticle forward-scattering phase shifts 
$\Phi_{\alpha\alpha'}(q,q')$. These are defined in 
Ref. \cite{Carmelo92}. In particular, the velocities 
$v_{\alpha}\equiv v_{\alpha}(q_{F\alpha}^{(+)})$
and the parameters 

\begin{equation}
\xi_{\alpha\alpha '}^j=\delta_{\alpha\alpha '}+ 
\Phi_{\alpha\alpha '}(q_{F\alpha}^{(+)},q_{F\alpha '}^{(+)})+
(-1)^j\Phi_{\alpha\alpha '}(q_{F\alpha}^{(+)},q_{F\alpha '}^{(-)})
\, , \hspace{1cm} j=0,1
\, ,
\end{equation}
play a determining role at the critical point. 
($\xi_{\alpha\alpha '}^1$ are the entries of the transpose
of the dressed-charge matrix \cite{Frahm}.) 

It is useful for the studies of this paper to present
here the different limiting values of the $\xi_{\alpha\alpha '}^1$
parameters (A4) (below $W$ is the Mott-Hubbard gap and
half filling is reached when $\mu\rightarrow W/2$):

For $U\rightarrow 0$ when $\mu >W /2$ and $H>0$ 

\begin{equation}
\xi^{1} = 
\left[\begin{array}{cc}
1
\hspace{0.5cm} & 
0\\
1
\hspace{0.5cm} & 
1
\end{array}
\right] \, .
\end{equation}

For $H\rightarrow 0$ when $\mu >W /2$ 

\begin{equation}
\xi^{1} = 
\left[\begin{array}{cc}
\xi_0
\hspace{0.5cm} & 
\xi_0/2\\
0
\hspace{0.5cm} & 
1/\sqrt{2}
\end{array}
\right] \, .
\end{equation}

For $H\rightarrow 0$ and $U\rightarrow 0$ when 
$\mu >W /2$

\begin{equation}
\xi^{1} = 
\left[\begin{array}{cc}
\sqrt{2}
\hspace{0.5cm} & 
1/\sqrt{2}\\
0
\hspace{0.5cm} & 
1/\sqrt{2}
\end{array}
\right] \, .
\end{equation}

For $H\rightarrow 0$ and $U\rightarrow\infty$ when 
$\mu >W /2$

\begin{equation}
\xi^{1} = 
\left[\begin{array}{cc}
1
\hspace{0.5cm} & 
1/2\\
0
\hspace{0.5cm} & 
1/\sqrt{2}
\end{array}
\right] \, .
\end{equation}

For $H\rightarrow H_c$

\begin{equation}
\xi^{1} = 
\left[\begin{array}{cc}
1
\hspace{0.5cm} & 
0\\
\eta_0
\hspace{0.5cm} & 
1
\end{array}
\right] \, ,
\end{equation}
(where $\eta_0={2\over {\pi}}\tan^{-1} (
{\sin (\pi n)\over u})$.) 

For $\mu\rightarrow W /2$ 

\begin{equation}
\xi^{1} = 
\left[\begin{array}{cc}
1
\hspace{0.5cm} & 
\xi^{1}_{cs}\\
0
\hspace{0.5cm} & 
\xi^{1}_{ss}
\end{array}
\right] \, .
\end{equation}

For $\mu\rightarrow W /2$ and $H\rightarrow 0$ 

\begin{equation}
\xi = 
\left[\begin{array}{cc}
1
\hspace{0.5cm} & 
1/2\\
0
\hspace{0.5cm} & 
1/\sqrt{2}
\end{array}
\right] \, .
\end{equation}

For $\mu\rightarrow W /2$ and $H\rightarrow H_ c$ 

\begin{equation}
\xi^{1} = 
\left[\begin{array}{cc}
1
\hspace{0.5cm} & 
0\\
0
\hspace{0.5cm} & 
1
\end{array}
\right] \, .
\end{equation}

For $\mu\rightarrow W /2+4t$ when $H>0$ 

\begin{equation}
\xi^{1} = 
\left[\begin{array}{cc}
1
\hspace{0.5cm} & 
0\\
0
\hspace{0.5cm} & 
1
\end{array}
\right] \, .
\end{equation}

For $H\rightarrow 0$ and $\mu\rightarrow W /2+4t$ 

\begin{equation}
\xi^{1} = 
\left[\begin{array}{cc}
1
\hspace{0.5cm} & 
1/2\\
0
\hspace{0.5cm} & 
1/\sqrt{2}
\end{array}
\right] \, .
\end{equation}

Further information on the pseudoparticle representation
can be found, for example, in Refs. 
\cite{Carmelo94,Carmelo92,Carmelo92b,Carmelo96}.


\newpage
\centerline{TABLE}
\vspace{2.5cm}
\narrowtext
\begin{tabbing}
\sl \hspace{1.3cm} \= \sl $U\rightarrow 0 \, (a)$ \hspace{0.3cm} \= 
\sl $U\rightarrow 0 \, (b)$ \hspace{0.3cm} \= \sl 
$n\rightarrow 0 \, (c)$ \hspace{0.3cm} \= \sl $m\rightarrow 0
\, (d)$ 
\hspace{0.3cm} \= \sl $n_{\downarrow}\rightarrow 0 \, (e)$ 
\hspace{0.7cm}\= \sl $n\rightarrow 1 \, (f)$
\\ $\hat{\rho}_{c\iota}(k)$ 
\> $\hat{\rho}_{\uparrow\iota}(k)$ 
\> ${\hat{\rho}_{\rho\iota}(k)\over{\sqrt{2}}}$ 
\> $\hat{\rho}_{\rho\iota}(k)$
\> $\xi^0_{cc}\hat{\rho}_{\rho\iota}(k)$
\> $\sum_{\sigma} {\cal U}_{\sigma c} 
   \hat{\rho}_{\sigma\iota}(k)$
\> $\hat{\rho}_{\rho\iota}(k)$\\ 
$\hat{\rho}_{s\iota}(k)$ 
\> $\hat{\rho}_{\downarrow\iota}(k)$ 
\> $-{\hat{\rho}_{\sigma_z\iota}(k)\over {\sqrt{2}}}$ 
\> $\hat{\rho}_{\downarrow\iota}(k)$
\> $-{\hat{\rho}_{\sigma_z\iota}(k)\over {\sqrt{2}}}$
\> $\hat{\rho}_{\downarrow\iota}(k)$ 
\> $\sum_{\sigma} {\cal U}_{\sigma s} 
   \hat{\rho}_{\sigma\iota}(k)$ 
\label{tableI}
\end{tabbing}
\vspace{0.5cm} 
TABLE -- Limiting values of the $c$ and $s$ generators $(16)$
at $k=\iota {2\pi\over {N_a}}$ and for (a) $U\rightarrow 0$ 
when $0<n<1$ and $n_{\uparrow}
>n_{\downarrow}$; (b) $U\rightarrow 0$ when $0<n<1$ and 
$n_{\uparrow}=n_{\downarrow}$; (c) $n\rightarrow 0$ when 
$n_{\uparrow}>n_{\downarrow}$ and $U>0$; (d) $m\rightarrow 0$ 
when $0<n<1$ and $U>0$; (e) $n_{\downarrow}\rightarrow 0$
when $n_{\downarrow}<n_{\uparrow}$ (here ${\cal U}_{\uparrow c}
=1$); and (f) $n\rightarrow 1$ when $U>0$.
\end{document}